\documentclass[journal=Langmuir,manuscript=article]{achemso}

\usepackage{chemformula} 
\usepackage[T1]{fontenc} 

\usepackage{color}
\usepackage[version=3]{mhchem}

\author{Mahtab Masouminia}
\affiliation[McMaster University]
{Department of Civil Engineering, McMaster University, Hamilton, Ontario L8S 4L8, Canada}
\author{Kari Dalnoki-Veress}
\affiliation[McMaster University]
{Department of Physics \& Astronomy, McMaster University, Hamilton, Ontario L8S 4L8, Canada}
\author{Benzhong Zhao}
\email{robinzhao@mcmaster.ca}
\affiliation[McMaster University]
{Department of Civil Engineering, McMaster University, Hamilton, Ontario L8S 4L8, Canada}

\title{Wettability alteration in thiolene-based polymer microfluidics:\\surface characterization and advanced fabrication techniques}

\abbreviations{IR,NMR,UV}

\begin{document}

%
%
%
%
%

\begin{abstract}
  Wettability plays a significant role in controlling multiphase flow in porous media for many industrial applications, including geologic carbon dioxide sequestration, enhanced oil recovery, and fuel cells. Microfluidics is a powerful tool to study the complexities of interfacial phenomena involved in multiphase flow in well-controlled geometries. Recently, the thiolene-based polymer called NOA81 emerged as an ideal material in the fabrication of microfluidic devices, since it combines the versatility of conventional soft photolithography with a wide range of achievable wettability conditions. Specifically, the wettability of NOA81 can be continuously tuned through exposure to high-energy UV. Despite its growing popularity, the exact physical and chemical mechanisms behind the wettability alteration have not been fully characterized. 
  
  Here, we apply different characterization techniques, including contact angle measurements, X-ray photoelectron spectroscopy (XPS), and atomic force microscopy (AFM) to investigate the impact of high-energy UV on the chemical and physical properties of NOA81 surfaces. We find that high-energy UV exposure increases the oxygen-containing polar functional groups, which enhances the surface energy and hydrophilicity of NOA81. Additionally, our AFM measurements show that spin-coated NOA81 surfaces have a roughness less than a nanometer, which is further reduced after high-energy UV irradiation. Lastly, we advance the state-of-the-art of NOA81-based microfluidic systems by creating i) a 2D surface with controlled wettability gradient and ii) a 3D column packed with monodisperse NOA81 beads of controlled size and wettability.
\end{abstract}

\section{Introduction}

Fluid-fluid displacement in small, confined geometries is strongly influenced by the relative affinity of the surrounding solid for the different fluids (i.e., wettability)~\cite{cueto-prl-2012,zhao-prl-2018,zhao-pnas-2019}. Wettability at the small-scale has important implications in a variety of large-scale natural and industrial processes, including vadose zone hydrology~\cite{debano-soilsci-1971,hill1972wetting,cueto-prl-2008}, enhanced oil recovery~\cite{morrow-jpt-2011,chen-spej-2013}, geologic carbon and hydrogen storage~\cite{chalbaud-awr-2009,iglauer-grl-2021}, and electrochemical energy conversion and storage~\cite{shrestha-jps-2018,zhao-crps-2021}. Microfluidic devices offer a powerful experimental platform to study the impact of wettability on fluid-fluid displacement, since they allow for direct visualization of the fluid interfaces and they can be fabricated with controllable geometries~\cite{sinton-labchip-2014,yun-labchip-2017,gupta-langmuir-2018,anbari-small-2018}. 

Different techniques have been introduced to tune the wettability condition of microfluidic experiments, which include the use of different fluid-fluid pairs~\cite{golmohammadi2021}, chemical vapor deposition (CVD) or liquid solution deposition of silane molecules~\cite{zhang-acsami-2016,silverio-colloids-2019}, oxidization of polydimethylsiloxane (PDMS) surfaces via either corona discharge~\cite{davies2012formation} or oxygen plasma~\cite{tan-biomicr-2010} treatment, and coating PDMS surfaces with a sol-gel layer that is functionalized with fluorinated and photoreactive silanes~\cite{abate-labchip-2008,abate-labchip-2010a}. Recently, the polymer called NOA81 (Norland Products, USA) has emerged as an alternative material for fabricating microfluidic devices with controlled wettability conditions~\cite{levache-labchip-2012,zhao-pnas-2016,odier-prl-2017}. NOA81 is a thiolene-based photocurable resin that enables patterning of submicron-size features via soft imprint lithography~\cite{bartolo-labchip-2008}. The wettability of NOA81 surfaces can be tuned via exposure to high-energy UV irradiation. In addition to the many positive attributes including solvent resistance, biocompatibility, and high elastic modulus~\cite{sollier-labchip-2011,geczy-labchip-2019a}, NOA81 offers the following advantages when it comes to wettability alteration: (i) its wettability can be continuously tuned and controlled by varying the duration of high-energy UV exposure~\cite{levache-labchip-2012}. Specifically, the wettability as measured by the contact angle $\theta$ of water in silicone oil varies over a wide range ($\theta=7^{\circ}$~--~$120^{\circ}$)~\cite{levache-prl-2014}. This wettability alteration is especially desirable since both water and silicone oil are commonly used and well characterized analog liquids in studies of capillarity and interfacial phenomena. (ii) the change in wettability is stable over a timescale of many days~\cite{levache-labchip-2012}.  

Despite the growing popularity of NOA81 in microfluidic studies (e.g.~\cite{gu-labchip-2010,wagli2011norland,xiao-langmuir-2018,moonen2018single,vavra-scirep-2020,bajgiran-analyst-2021}), the physical and chemical mechanisms behind the UV-induced wettability alteration have not been characterized. Our work aims to fill this knowledge gap. To this end, we first create a highly uniform NOA81 thin film via spin coating on a silicon wafer. We then employ characterization techniques including contact angle, X-ray photoelectron spectroscopy (XPS) and atomic force microscopy (AFM) measurements to investigate changes to the NOA81 surface as a result of high-energy UV irradiation. Our analyses show that wettability alteration on NOA81 surfaces arise as a result of the emergence of polar, oxygen-containing functional groups after high-energy UV exposure. Finally, we extend the potential use cases for NOA81 by introducing procedures to generate (i) NOA81 surfaces with controlled wettability gradients and (ii) monodisperse NOA81 beads with controlled size and wettability.

\section{Surface fabrication}

\begin{figure} [!h]
	\centering
	\includegraphics[width=9cm]{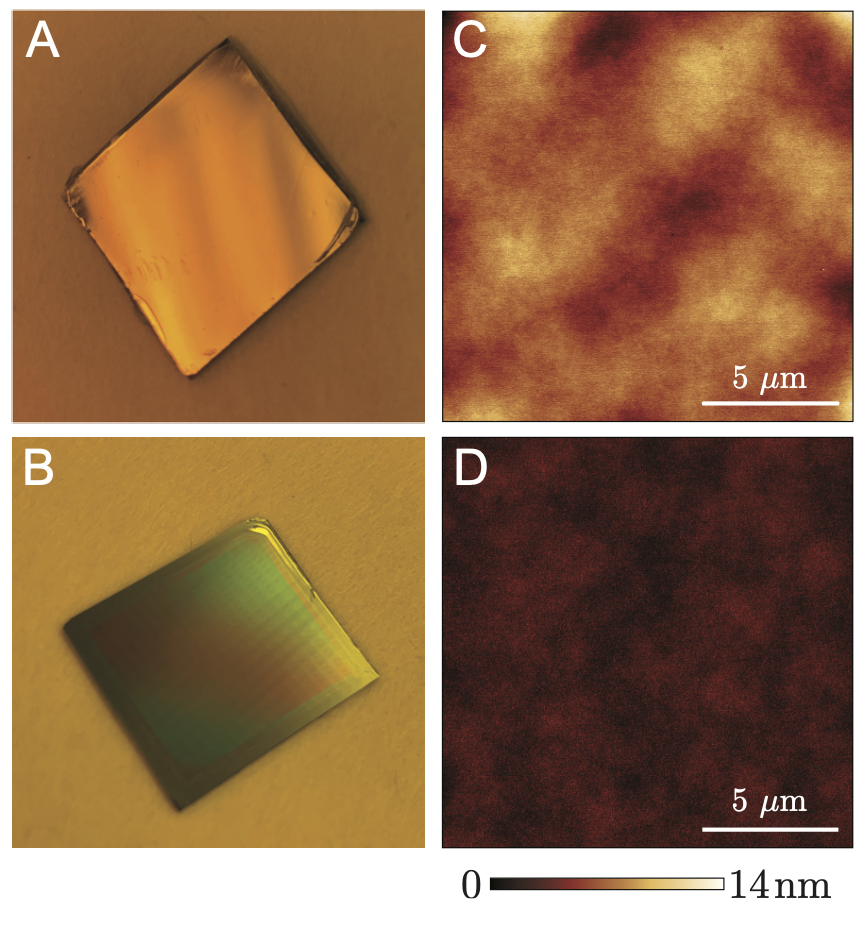}
	\caption{(A) Thin film of NOA81 cured on an $1''\times1''$ silicon wafer created by PDMS replica molding. The thickness of the NOA81 is 100$\;\mu\mbox{m}$. (B) Thin film of NOA81 cured on an $1''\times1''$ silicon wafer created by spin coating. The thickness of the NOA81 is $\sim$7.5$\;\mu\mbox{m}$. (C) AFM surface roughness measurements of the NOA81 film created by PDMS replica molding. (D) AFM surface roughness measurements of the NOA81 film created by spin coating. Both methods generate smooth coatings with roughness less than 14$\;\mbox{nm}$, though spin coating attains a significantly more homogeneous film.~\label{fig:spincoat}}
\end{figure}

A survey of the literature shows that existing NOA81 surfaces in microfluidics applications are fabricated via replica molding. In this method, NOA81 is sandwiched between a flat surface and a mold (typically made of PDMS) in the negative shape of the desired micro-pattern before the NOA81 is cured~\cite{bartolo-labchip-2008,levache-labchip-2012,levache-prl-2014,zhao-pnas-2016,odier-prl-2017}. To generate a flat NOA81 surface via replica molding, we first fabricate a flat PDMS (Sylgard 184, Dow Corning, USA) substrate that is cured on a silicon wafer. We then sandwich a drop of NOA81 between a silicon wafer and the flat PDMS substrate, separated by 100~$\mu$m thick precision shims. After curing the NOA81 with 365~nm UV light for 10~s, we peel off the PDMS substrate to reveal the NOA81 surface~(Fig.~\ref{fig:spincoat}A). We expose the NOA81 surface to UV for an additional 20~s after PDMS removal to cure the ultra thin superficial layer of NOA81~\cite{bartolo-labchip-2008}. We measure the topography and roughness of the NOA81 surface with an atomic force microscope (MultiMode 8-HR, Bruker, USA) with a scan area of $15\times15~\mu${m}$^2$. We find that the NOA81 surface made by replica molding has a root mean square (RMS) roughness of $\sim$1.32~nm, with a correlation length of $\sim$0.95~$\mu$m~(Fig.~\ref{fig:spincoat}C).

We then create a thin film of NOA81 on a silicon wafer by spin-coating. The silicon wafer is first treated in an oxygen plasma asher for 1~min (Model PT7150, Bio-Rad, USA)~---~NOA81 thin films spin-coated on untreated silicon wafers are unstable at ambient temperature ($20~^{\circ}$C) and they undergo spinodal dewetting. We achieve a stable NOA81 thin film at spin speed of 4000~rpm for 30~s, which is then cured with 365~nm UV light for 10~s~(Fig.~\ref{fig:spincoat}B). The thickness of the cured NOA81 thin film is $\sim$7.5~$\mu$m as measured by ellipsometry (Model M-2000, J. A. Woolam, USA). AFM measurement of the spin-coated NOA81 surface reveals that it has an RMS roughness of $\sim$0.59~nm, with a correlation length of $\sim$0.6~$\mu$m~(Fig.~\ref{fig:spincoat}D). Therefore, spin-coating generates a smoother NOA81 surface at the nanoscale compared to replica molding, and we use the spin-coated NOA81 surface in charaterizations discussed in subsequent sections.   

\section{Surface property change due to high-energy UV exposure}

\subsection{Contact angle characterization}

\begin{figure} 
	\centering
	\includegraphics[width=10cm]{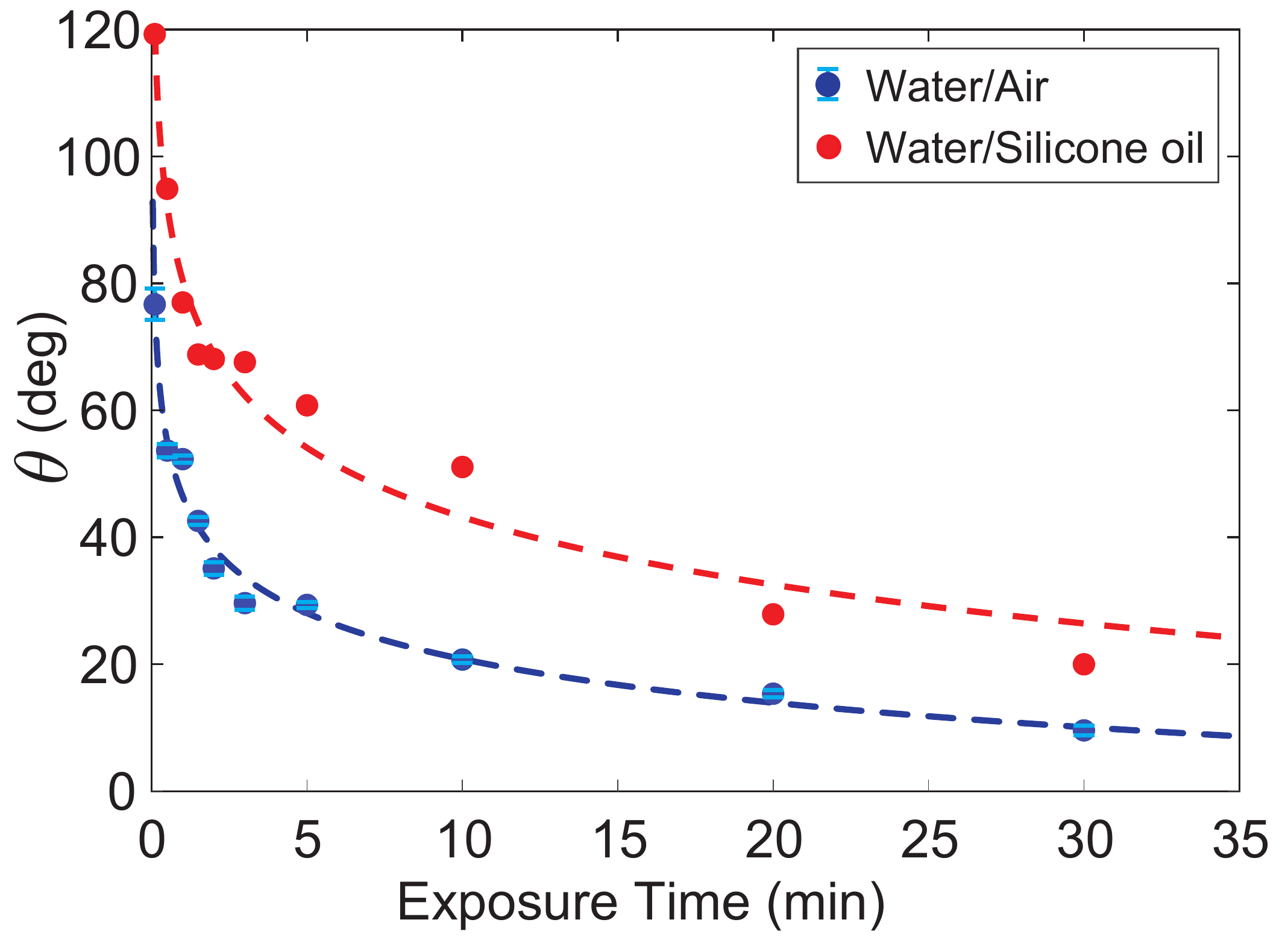}
	\caption{Static advancing contact angle $\theta$ of water on spin-coated NOA81 surfaces as a function of high-energy UV exposure time for two different ambient fluids: air (blue circles) and silicone oil (red circles). The contact angle of the NOA81 surface decreases with increasing high-energy UV exposure time in the presence of both ambient fluids. The dashed lines represent the power fit (Eq.~\ref{eq:theta_fit}).} \label{fig:contactangle}
\end{figure}

NOA81 surface is known to become more hydrophilic after exposure to high-energy UV irradiation. \citet{levache-labchip-2012} characterized the UV-induced contact angle change in NOA81 surface made by replica molding for water-in-air and water-in-hexadecane oil systems. Here, we treat the spin-coated NOA81 surfaces with high-energy UV for different durations in a UV-ozone cleaner (Model T0606B, UVOCS, USA), which generates UV emissions in the 185 and 254~nm range. We characterize the wettability of the surfaces by measuring the static advancing contact angle of water using a contact angle goniometer immediately after high-energy UV exposure. The measurements are conducted for both water-in-air and water-in-silicone oil (Millipore Sigma, USA) systems. The water-in-air contact angle measurements are repeated at six different spots on each sample. We observe remarkable spatial uniformity in contact angle, which is illustrated by maximal dispersions of $\sim2^{\circ}$~(Fig.~\ref{fig:contactangle}).

The change in NOA81 surface wettability as a function of high-energy UV exposure can be described by a simple power law fit:

\begin{equation} \label{eq:theta_fit} 
  \theta(t)=a+bt^{-c},
\end{equation}

where $\theta$ is the static advancing contact angle of water, $t$ is the exposure duration in minutes, $a,b,c$ are fitting parameters. We find $a=-112,b=158,c=0.0767$ for the water-in-air system; $a=-398,b=478,c=0.0352$ for the water-in-silicone oil system.



\subsection{Surface chemistry characterization}

\begin{figure} [!h]
	\centering
	\centering
	\includegraphics[width=12cm]{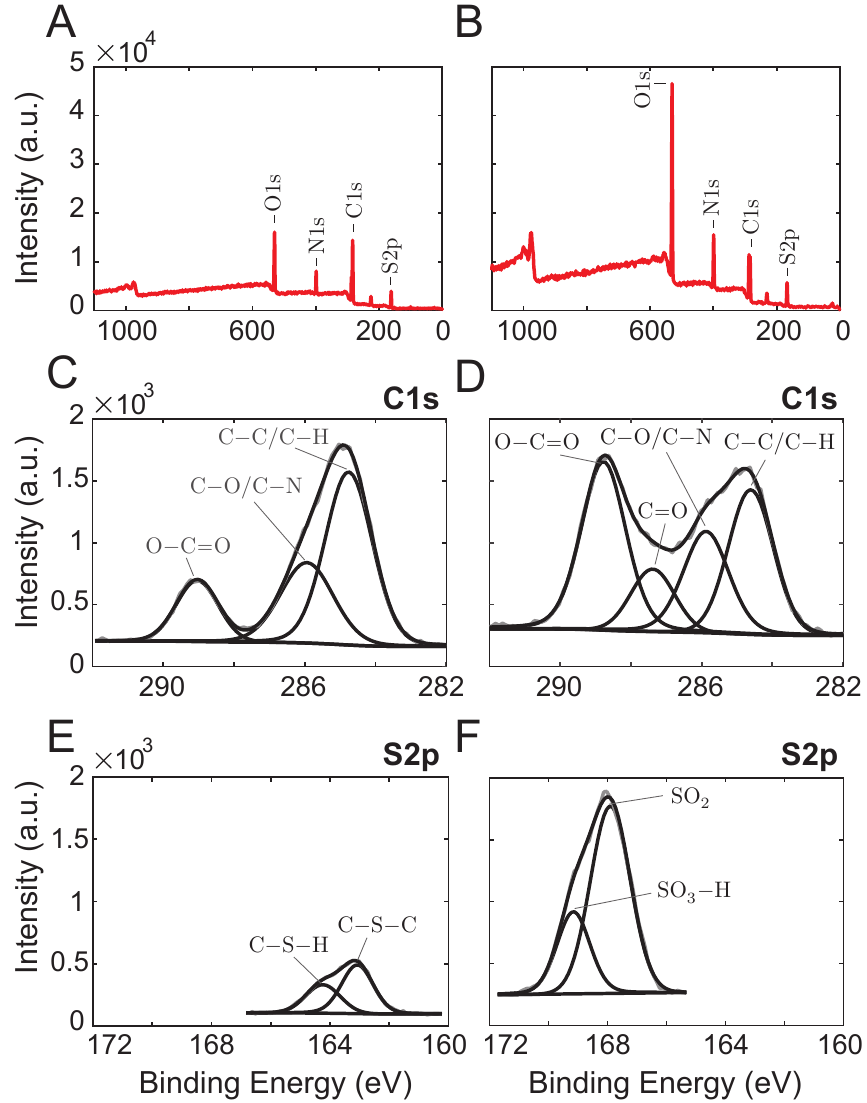}
	\caption{We characterize the surface chemical compositions of the untreated and UV-treated NOA81 films via XPS. The spectra show the average of the scans performed per sample at three different points. (A) full spectrum, (C) C1s spectra, and (E) S2p spectra of the untreated NOA81 surface (left column). (B) full spectrum, (D) C1s spectra, and (F) S2p spectra of the NOA81 surface that has been exposed to high-energy UV irradiation for 30 min (right column). The XPS survey demonstrates that increases in carbonyl (\ce{C=O}), carboxyl (\ce{O-C=O}), sulfone (\ce{SO2}), sulfonate (\ce{SO3-H}), ether (\ce{C-O}), and amine (\ce{C-N}) polar functional groups and decreases in hydrocarbon (\ce{C-C}/\ce{C-H}) and thiol (\ce{C-S=H}) non-polar functional groups in the UV-treated sample contribute to its surface hydrophilicity. \label{fig:XPSspectra}}
\end{figure}

The dramatic change in NOA81's wettability after high-energy UV exposure indicates a corresponding change in its surface chemistry. We characterize the surface chemistry of NOA81 surfaces with XPS, which provides quantitative information on the elemental composition, as well as chemical composition of the surface~\cite{watts2019xps}. Our XPS analyzer (Model PHI Quantera II, Physical Electronics, USA) is equipped with a monochromatic Al K$\alpha$ X-ray (1486.6 eV) source and operated under ultra-high vacuum ($\sim 10^{-9}$ Torr) at a photoelectron take-off angle of 45$^{\circ}$. We use a pass energy of 224 eV with an energy step of 0.8 eV to obtain the full spectrum (Fig.~\ref{fig:XPSspectra}A, B), which includes all elements present on the sample surface. Since carbon and sulfur are major structural elements in thiolene-based polymers, we additionally use a pass energy of 23.5 eV with an energy step of 0.1 eV to acquire higher resolution data of the carbon 1s (C1s) and sulfur 2p (S2p) spectra (Fig.~\ref{fig:XPSspectra}C-F). All spectra are calibrated with reference to the C1s spectrum at 284.8 eV and analyzed using Gaussian-Lorentzian peak deconvolution along with Shirley background reduction using MultiPak Spectrum software~\cite{luthin2000carbon}.

The most striking difference between XPS full spectrum scans of untreated and treated NOA81 surfaces is the sharp rise in the peak of oxygen 1s (O1s) region, which demonstrates a significant increase in the number of oxygen-containing functional groups on the NOA81 surface after high-energy UV irradiation~(Fig.~\ref{fig:XPSspectra}A, B). The higher O1s peak of the treated NOA81 sample indicates an increase in the polarity, and hence hydrophilicity of the surface, since oxygen lone pairs are strong hydrogen bond acceptors. 

The C1s spectrum of the untreated NOA81 surface has three peaks at binding energies of 284.7 eV, 285.9 eV, and 289.0 eV correspond to hydrocarbon (\ce{C-C}/\ce{C-H}) non-polar groups, ether/amine (\ce{C-O}/\ce{C-N}), and carboxyl (\ce{O-C=O}) polar functional groups (Fig.~\ref{fig:XPSspectra}C). After high-energy UV treatment, we observe a significant increase in the peak of carboxyl functional group and decrease in the peak of hydrocarbon functional group, in addition to the emergence of carbonyl (\ce{C=O}) polar functional group at a binding energy of 287.4 eV (Fig.~\ref{fig:XPSspectra}D)~\citep{xie-rsc-2021}. The S2p spectrum of the untreated NOA81 surface  shows two peaks at binding energies of 163.1 eV and 164.2 eV, which correspond to sulfide (\ce{C-S-C}) slightly polar functional group and thiol (\ce{C-S=H}) non-polar functional group (Fig.~\ref{fig:XPSspectra}E). After high-energy UV treatment, we observe the emergence of two oxidized sulfur species on the S2p spectrum, which correspond to sulfone (\ce{SO2}) at a binding energy of 167.9 eV and sulfonate (\ce{SO3-H}) at a binding energy of 169.1 eV (Fig.~\ref{fig:XPSspectra}F). Both sulfone and sulfonate are polar functional groups~\citep{lindberg1970molecular,gobbo2013facile}. The increase in polar functional groups and decrease in non-polar functional groups on the UV-treated sample contribute to surface hydrophilicity~\citep{mathieson-adh-1996,yang-chemrev-2013,roy2010surface,ren-appsurf-2017,vo2009polymers,poulsson-langmuir-2009attachment}.  

In addition to the XPS spectra scans, we measure the atomic concentrations of carbon (C), oxygen (O), nitrogen (N), and sulfur (S) on NOA81 surfaces that have been exposed to high-energy UV irradiation for different durations (Table~\ref{tbl:content}). We find that the ratios of oxygen to carbon atoms (O/C) and nitrogen to carbon atoms (N/C) increase monotonically with increasing exposure time, which is indicative of rising polar functional groups and reflected in the increasing hydrophilicity ~\citep{lehocky-cjp-2006,sharma2007stability}.

\begin{table} [h]
	\caption{Atomic concentrations of NOA81 surfaces for different high-energy UV treatment times. The atomic concentration values are the average concentrations collected at three different points on the surface. The XPS analysis shows a decrease in carbon content and an increase in oxygen content as well as atomic O/C and N/C ratios, which attribute to the hydrophilicity of the UV-treated sample.}
	\label{tbl:content}
	\begin{tabular*}{0.6\textwidth}{@{\extracolsep{\fill}}ccccccc}
		\hline
		\textbf{Exposure} & \textbf{C} & \textbf{O} & \textbf{N} & \textbf{S} & \textbf{O/C} & \textbf{N/C} \\
		\textbf{[min]} & \textbf{[\%]} & \textbf{[\%]} & \textbf{[\%]} & \textbf{[\%]} & \textbf{ratio} & \textbf{ratio} \\
		\hline
		0 & 63.9 & 21.9 & 8.9 & 5.3 & 0.34 & 0.14 \\
		1 & 56.9 & 29.4 & 8.6 & 5.1 & 0.52 & 0.15 \\
		5 & 51.5 & 33.9 & 10.1 & 4.5 & 0.66 & 0.19 \\
		10 & 50.5 & 34.9 & 10.2 & 4.4 & 0.69 & 0.20 \\
		30 & 40.7 & 39.2 & 16.0 & 4.1 & 0.96 & 0.39 \\
		\hline
	\end{tabular*}
\end{table}

\subsection{Surface topology characterization}

\begin{figure}
	\centering
	\includegraphics[width=9cm]{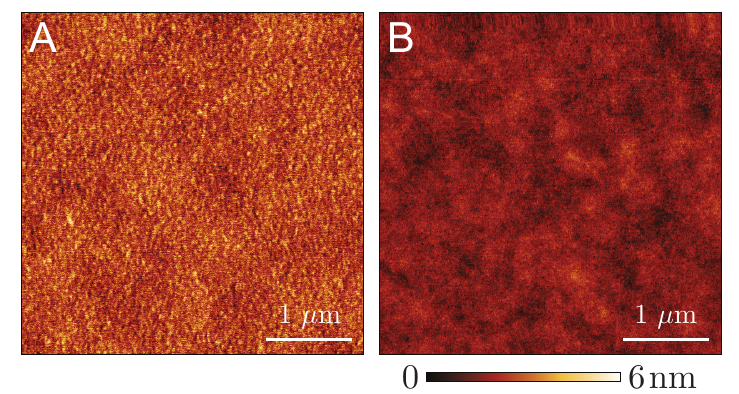}
	\caption{AFM surface morphologies of the NOA81 films (A) not exposed to the high-energy UV and (B) exposed to the high-energy UV for 30 min demonstrate that high-energy UV exposure reduces the roughness of the NOA81 film. \label{fig:AFM}} 
\end{figure}

We characterize the physical changes to the NOA81 surface as a result of high-energy UV irradiation. Specifically, AFM measurements demonstrate that high-energy UV exposure makes NOA81 surfaces smoother, though the effect is quite small (Fig.~\ref{fig:AFM}). The untreated NOA81 has an RMS roughness of 0.61~nm with a correlation length of 19.47~nm, while the NOA81 surface that has been exposed to high-energy UV for 30 min has an RMS of 0.37~nm and a correlation length 63.23~nm. Additionally, we measure the thickness of the NOA81 film before and after high-energy UV irradiation using an ellipsometer. Our measurements show that the NOA81 film thickness decreases from $7.5~\mu$m to $7.2~\mu$m after being exposed to high-energy UV for 30 min.


\section{Extension of use cases}

\subsection{Surface with controlled wettability gradient}

Wettability gradient is a common occurrence in many natural and industrial processes---for example, the diffusion of biosurfactants produced by bacteria can locally make soil particles more water-wet~\cite{yang2021}, and the dispersion of asphaltenes can locally make reservoir rocks more oil-wet~\cite{kaminsky-spe-1997}. In the context of lab-on-a-chip technology, wettability gradients have been extensively utilized in the past two decades for the spontaneous and directional transport of liquid droplets in analytical chemistry and bioassay applications~\cite{maria-scirep-2017,liu-scirep-2017,qi-labchip-2019}. 

\begin{figure}
	\centering
	\includegraphics[width=16cm]{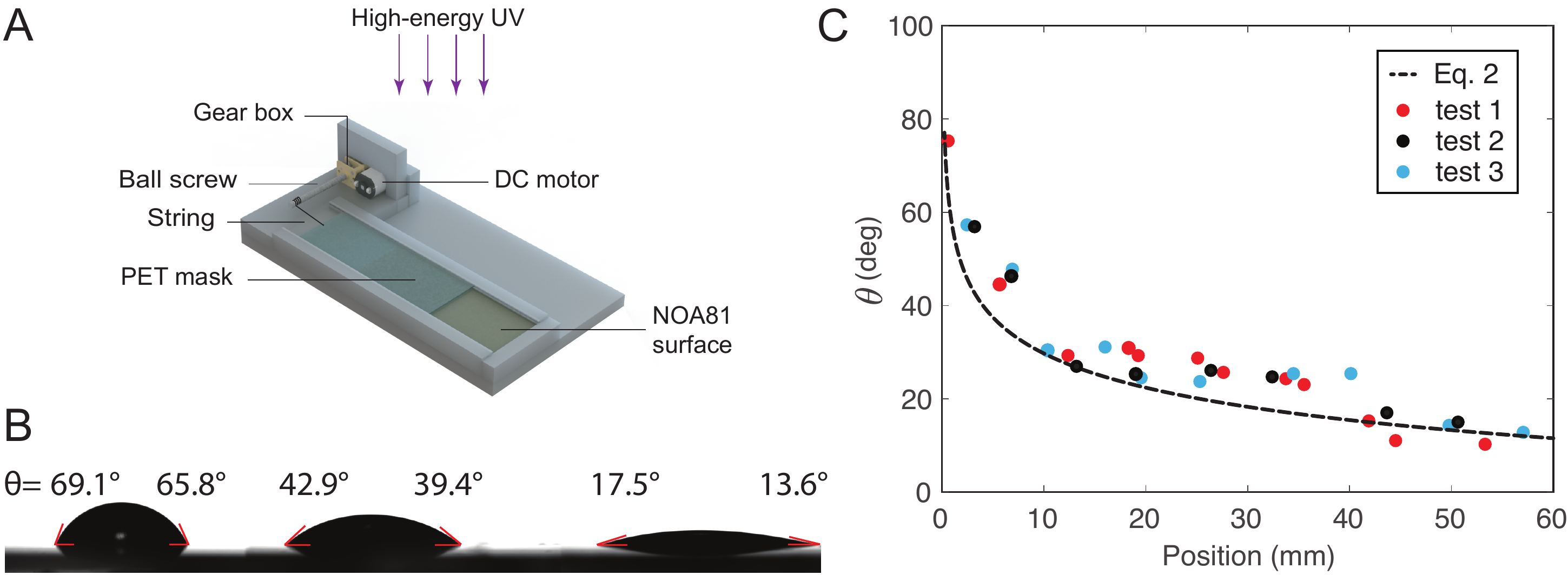}
	\caption{(A) Schematic of the fabrication platform to generate wettability gradients on NOA81 surfaces. We control the amount of high-energy UV exposure on the NOA81 surface via an overlying UV-blocking PET mask that is gradually retracted by a DC motor. (B) Left and right contact angles of three water droplets placed along a NOA81 surface with a wettability gradient in the presence of air. (C) The measured contact angles along the wettability gradient surface (circles) agrees with predictions of Eq.~\ref{eq:gradient} (dashed line). \label{fig:gradient}}
\end{figure}

Here, we introduce a fabrication technique to generate wettability gradients on NOA81 surfaces. Specifically, we take advantage of the property that NOA81 becomes more hydrophilic with increasing exposure to high-energy UV irradiation (Fig.~\ref{fig:contactangle}) via the application of a UV-blocking polyethylene terephthalate (PET) mask. The PET mask is placed on top of the NOA81 surface and attached to a direct current (DC) motor via a taut string on one side. The motor gradually pulls back the mask along a track at a constant velocity, exposing the underlying NOA81 surface to high-energy UV (Fig.~\ref{fig:gradient}A). The cumulative exposure of the NOA81 surface varies along the travel direction of the mask, which results in a spatial wettability gradient:  

\begin{equation} \label{eq:gradient} 
  \theta(x)=a+bv^cx^{-c},
\end{equation} 

where $a,b,c$ are the fitting parameters described in Eq.~\ref{eq:theta_fit}, $v$ is the velocity of the PET mask, and $x\in{[0,L]}$ is the position along the substrate of length $L$. We generate a wettability gradient on an $L=60$~mm-long NOA81-coated microscope slide with $v=2.4$~mm/min (Fig.~\ref{fig:gradient}), which agrees well with the contact angle distribution predicted by Eq.~\ref{eq:gradient}. We note that one can design and achieve different types of wettability gradients by employing a varying mask velocity in time, which can be accomplished by connecting the DC motor to a programmable microcontroller (e.g., Arduino Nano).

\subsection{Spherical beads with controlled wettability and size}

NOA81 micromodels in the existing literature have been constructed in Hele-Shaw-type geometries (e.g.,~\cite{zhao-pnas-2016, xiao-langmuir-2018,vavra-scirep-2020,bajgiran-analyst-2021}), where fluid flow takes place between two parallel flat surfaces separated by a small gap. An alternative method of creating micromodels is by packing spherical beads between parallel plates~\cite{maloy-prl-1992,maloy-prl-1985,holtzman-prl-2012,trojer-prapp-2015} or in a tube~\cite{datta-prl-2013,datta-physfluids-2014,singh-scirep-2017}. This type of micromodel has made fundamental contributions to our understanding of multiphase flow in porous media. 

\begin{figure} [!h]
	\centering
	\includegraphics[width=16cm]{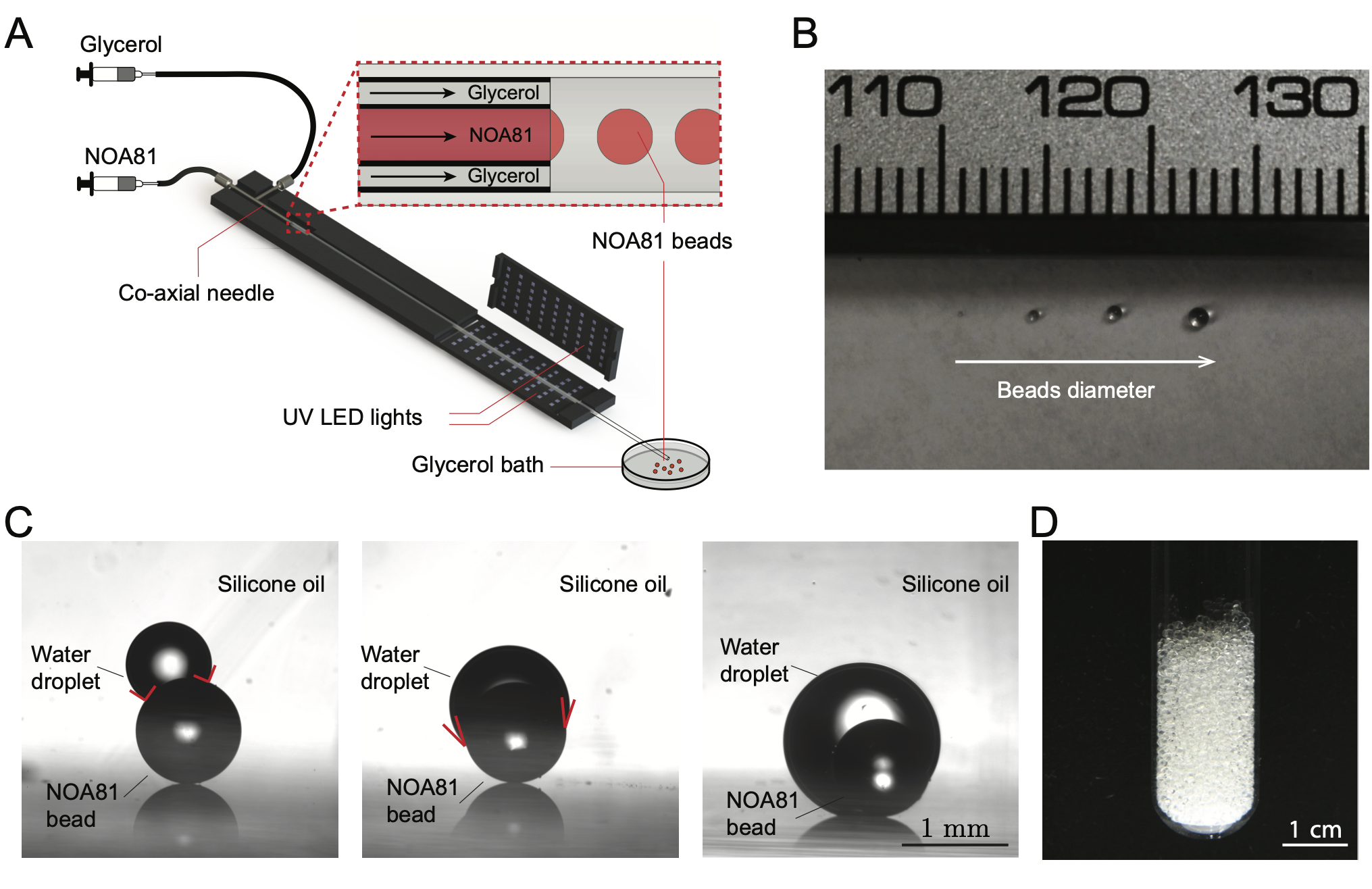}
	\caption{(A) Schematic of the microfluidic platform for generating NOA81 sphere beads with controlled size and wettability. We use a coaxial needle as a droplet generator to form the NOA81 droplets. Glycerol is the continuous phase, and NOA81 is the dispersed phase. UV LED lights downstream cure the NOA81 droplets. (B) Pictures of NOA81 beads with different diameters (220, 500, 715, and 930 $\mu$m from left to right). (C) Left: The contact angle of water on an untreated NOA81 bead in a bath of silicone oil is $92^{\circ}$. Middle: The contact angle of water in silicone oil on a NOA81 bead that has been exposed to 30-min of high-energy UV irradiation is $26^{\circ}$. Right: The NOA81 bead that has been exposed to 2 hours of high-energy UV irradiation becomes fully encapsulated by the water drop. (D) A column packed with monodisperse NOA81 beads.}
	\label{fig:beads}
\end{figure}

Here, we introduce a simple microfluidic platform for generating NOA81 beads with controlled size and wettability. Upstream of the microfluidic platform consists of a coaxial needle (ramé-hart, USA) connected to a Tygon tubing. We operate the coaxial needle in a co-flow manner by injecting NOA81 through the smaller inner needle and glycerol (Sigma-Aldrich, USA) through the larger outer needle. In this setup, the less viscous NOA81 ($\mu_\text{NOA81}=300$~mPa$\cdot$s) forms a jet that penetrates into the more viscous glycerol ($\mu_\textrm{glycerol}=648$~mPa$\cdot$s). The jet of NOA81 becomes unstable and breaks into drops due to the Rayleigh-Plateau instability (Fig.~\ref{fig:beads}A and Movie S1)~\cite{utada-prl-2007,duncanson-labchip-2012}. The drop size is controlled by the injection rates of NOA81 and glycerol, and by the geometry of the coaxial needle. We use two coaxial needle models in our experiments~---~one model has an inner needle with inner diameter $\text{ID}=406~\mu$m and outer diamter $\text{OD}=711~\mu$m, external needle with $\text{ID}=840~\mu\text{m},\text{OD}=1240~\mu$m; while the other model has an inner needle with $\text{ID}=140~\mu\text{m},\text{OD}=305~\mu$m, and outer needle with $\text{ID}=584~\mu\text{m},\text{OD}=889~\mu$m. We achieve monodisperse NOA81 beads whose diameters range between $220$ to $930~\mu$m (Fig.~\ref{fig:beads}B). Downstream of the microfluidic platform consists of a chamber lined with UV LED lights, which cure the NOA81 droplets as they pass through. The cured NOA81 beads are collected, rinsed with deionized water, and dried with nitrogen gas. 

We characterize the wettability of the NOA81 beads by placing a small drop of water on the bead in a chamber filled with silicone oil, which yields a contact angle of 92$^{\circ}$ (Fig.~\ref{fig:beads}C and Movie S2). To make the NOA81 beads more hydrophilic, we expose them to high-energy UV irradiation in a quartz glass petri dish, since quartz glass has high transmission of the 185 and 254~nm UV light needed for wettability alteration. We attach a mirror to the bottom of the petri dish, which reflects incident UV rays back towards the bottom of the beads. We wrap the sides of the petri dish with reflective aluminum foil to further increase exposure uniformity. The contact angle of water in silicone oil on a NOA81 bead that has been exposed to high-energy UV for $30$~min is $26^{\circ}$ (Fig.~\ref{fig:beads}C). Interestingly, the water drop is able to pick up the UV-treated NOA81 bead due to capillary adhesion (Movie S3). Further increasing the exposure time and the water drop volume triggers the total encapsulation of the NOA81 bead by the water (Fig.~\ref{fig:beads}C and Movie S4). Finally, we demonstrate the scalability of the microfluidic platform by creating a column packed with NOA81 beads (Fig.~\ref{fig:beads}D), which could be used to investigate the impact of wettability and mixed-wettability on multiphase flow in truly 3D geometries.

\section{Conclusions}

We have presented a comprehensive investigation of the mechanisms behind wettability alteration on NOA81 surfaces as a result of high-energy UV irradiation. NOA81 is a thiolene-based photocurable polymer with growing popularity in microfluidics, since the wettability of NOA81 can be precisely tuned over a wide range of contact angles (Fig.~\ref{fig:contactangle}). We characterize the surface chemistry of NOA81 via XPS measurements, which reveal the abundance of hydrophobic non-polar functional groups (i.e., hydrocarbon and thiol) on untreated NOA81 surfaces, and the formation of hydrophilic polar functional groups (i.e., carboxyl, carbonyl, sulfone and sulfonate) on UV-treated NOA81 surfaces~(Fig.~\ref{fig:XPSspectra}). Our AFM measurements reveal the highly smooth nature of spin-coated NOA81 surfaces (Fig.~\ref{fig:spincoat}), whose roughness further decreases after exposure to high-energy UV irradiation (Fig.~\ref{fig:AFM}). Lastly, we demonstrate the ever-expanding potential of NOA81 in even more microfluidics applications by designing fabrication systems to generate i) a 2D surface with controlled wettability gradient (Fig.~\ref{fig:gradient}) and ii) monodisperse NOA81 beads with controlled size and wettability (Fig.~\ref{fig:beads}). Our work opens new avenues for studying the complex wettability-controlled interfacial phenomena in multiphase flow in porous media.

\begin{acknowledgement}

The authors thank Doris Stevanovic at the McMaster Center for Emerging Device Technologies (CEDT) for her training and guidance on clean room equipment. This reserach was supported by the Natural Sciences and Engineering Research Council of Canada (NSERC) Discovery Grants.

\end{acknowledgement}

\begin{suppinfo}

The following files are available free of charge.
\begin{itemize}
  \item movie\_S1.mp4: Video showing the generation of NOA81 droplets using a coaxial needle. Glycerol is the continuous phase and NOA81 is the dispersed phase.
  \item movie\_S2.mp4: Video showing the contact angle measurement of water on an untreated NOA81 bead in a bath of silicone oil.
  \item movie\_S3.mp4: Video showing the contact angle measurement of water in silicone oil on a NOA81 bead that has been exposed to 30-min of high energy UV irradiation.
  \item movie\_S4.mp4: Video showing the total encapsulation of a NOA81 bead that has been exposed to 2-hrs of high energy UV irradiation.
\end{itemize}

\end{suppinfo}

\bibliography{noa}

\end{document}